\RequirePackage[2020-02-02]{latexrelease}
\documentclass[reprint,
groupedaddress,
unsortedaddress,
 amsmath,amssymb, aps,
prb,rmp,prstab,prstper,floatfix]{revtex4-2}

\usepackage{graphicx}
\usepackage{xcolor}
\usepackage{multirow}
\usepackage{dcolumn}
\usepackage{bm}
\usepackage{hyperref}
\usepackage[mathlines]{lineno}
\usepackage{soul} 
\usepackage{comment}

\newcommand{\kk}{{\bf k}}

\newcommand{\ggf}{\mathrm{G}_0\mathrm{W}_0^{GF}}
\newcommand{\gac}{\mathrm{G}_0\mathrm{W}_0^{AC}}
\newcommand{\gowo}{\mathrm{G}_0\mathrm{W}_0}
\newcommand{\gw}{\mathrm{G}\mathrm{W}}
\newcommand{\scgw}{\mathrm{scG}\mathrm{W}}

\usepackage{hyperref}

\hypersetup{
    colorlinks=true,
    linkcolor=red,
    filecolor=magenta,      
    urlcolor=cyan,
}

\begin{document}


\title{GW space-time method: Energy band-gap of solid hydrogen}

\author{Sam Azadi}%
\email{s.azadi@imperial.ac.uk}
\affiliation{Department of Physics, King's College London, Strand, London WC2R 2LS,}
\affiliation{Department of Physics, Imperial College London, Exhibition Road, London SW7 2AZ, UK}
\author{Arkadiy Davydov}%
\email{adavydov@physik.uzh.ch}
\affiliation{Department of Physics, King's College London, Strand, London WC2R 2LS, UK}
\affiliation{Physik-Institut, University of Zurich, 8057 Zurich, Switzerland}
\affiliation{Scientific Computing Research Technology Platform, University of Warwick, Zeeman Building, CV4 7AL, Coventry, UK}
\author{Evgeny Kozik}
\affiliation{Department of Physics,  King's College London, Strand, London WC2R 2LS, UK}%

\date{\today}
\begin{abstract}
We implement the GW space-time method at finite temperatures, in which the Green’s function G and the screened Coulomb interaction W are represented in the real space on a suitable mesh and in imaginary time in terms of Chebyshev polynomials, paying particular attention to controlling systematic errors of the representation. Having validated the technique by the canonical application to silicon and germanium, we apply it to calculation of band gaps in hexagonal solid hydrogen with the bare Green’s function obtained from density functional approximation and the interaction screened within the random phase approximation (RPA). The band gap results, obtained from the asymptotic decay of the full Green’s function without resorting to analytic continuation, suggest that the solid hydrogen above 270 GPa can not adopt the hexagonal-closed-pack ($hcp$) structure. The demonstrated ability of the method to store the full G and W functions in memory with sufficient accuracy is crucial for its subsequent extensions to include higher orders of the diagrammatic series by means of diagrammatic Monte Carlo algorithms.
\end{abstract}


\maketitle


\section{\label{Intro} Introduction}
The metallization of solid hydrogen at low temperatures under extremely high pressures is an open challenging problem in condensed matter physics and materials science. It has been a subject of intensive studies since 1935~\cite{1935}---by experiment, theory and, more recently, computational techniques---stimulated, in particular, by the possibility that metallic hydrogen is a room-temperature superconductor, driven by conventional phonon-mediated mechanisms~\cite{Ashcroft}, harbors a new phase of matter in the low-temperature liquid state~\cite{Babaev, Bonev}, and can be found inside giant planets~\cite{Hemley,Ginzburg, RevMod12}. Nonetheless, determining the metallization mechanism and metallization pressure of solid hydrogen is still an open problem. 

Up to pressures of 430 GPa, around or beyond which the metallic state is expected to emerge, the detection of vibron excitations evidences that solid hydrogen is in a molecular form~\cite{Loubeyre20}. Thus, metallization is likely to emerge with compression either by a continuous closure of the energy band gap in a molecular crystal, or by a structural phase transition from the molecular insulator to an atomic or molecular metal. However, because determining the band gap experimentally under extreme conditions is very difficult, the available data is limited~\cite{Loubeyre20}. From the theoretical perspective, which metallization scenario is realized largely depends on the specific crystal structure assumed in the calculation. At the same time, despite an extensive amount of work by experiment~\cite{Hazen,Eremets,Howie,Howie2,Goncharov19,Howie3,Simpson,Dias,Liu,Eremets19,Loubeyre20} and theory~\cite{Pickard,Pickard2,Goncharov,Magdau,Naumov,Morales2013,Clay,Clay16,JETP,PRB13,PCCP17,Monserrat18,NJP,prl,Neil15,PRB17,PRB19,Ackland20}, there is no definitive conclusion on the crystal structure of solid hydrogen in the range of pressures where any of the metallization scenarios might take place. It has been recently observed in X-ray diffraction measurements~\cite{CJi} that, at least up to 250 GPa, solid hydrogen remains in the $hcp$ crystal form, albeit possibly with an increasing under compression anisotropy. It is therefore important to evaluate the pressure at which solid hydrogen becomes metallic by band gap closure, provided it remains an $hcp$ crystal. This is the problem addressed by our work. 

With the lack of direct experimental access, advanced computational techniques play a crucial role in elucidating the behavior of the band gap under compression. The diffusion quantum Monte Carlo (DMC) and variational quantum Monte Carlo (VMC) methods can be efficiently used to obtain the excitation energies above the ground state in a finite-size (FS) system, but the extrapolation to the thermodynamic limit is typically challenging~\cite{GCTAV, YY2020}. DMC calculations of excitation properties suffer from the simulation cell finite-size ($1/L$) errors and the $1/N$ effect: the variation in the total ground state energy due to the presence of a one- or two-particle excitations is inversely proportional to the number of electrons $N$ in the simulation cell. The large number of atoms per primitive cell in almost all of the predicted for solid hydrogen structures makes controlling the FS errors especially difficult.

An efficient approach to describing excitation properties directly in the thermodynamic limit is provided by the many-body perturbation theory in terms of Feynman diagrams. In this framework, the lowest-order in terms of the Green's function $G$ and screened Coulomb interaction $W$ GW approximation has been most widely used~\cite{Onida02, Hybertsen, Aryas}. The accuracy of the GW approach, assessed by comparison of the results with the experiment, can vary with the kind of approximation applied to $G$ and $W$ themselves. The natural self-consistent approach~\cite{Surh, Luo}---in which $G$ is determined by the self-energy in terms of $G$, $\Sigma = -GW$, with $W$ screened by the polarization given by $GG$---is known to overestimate the valence bandwidth of the Ge and Si crystals~\cite{Holm,Ku}. This discrepancy can be efficiently addressed~\cite{Markprl06, Kotani}, even in more challenging cases, such as MnO and NiO, by the quasiparticle self-consistent GW approach~\cite{Faleev04, Faleev06} (QSGW), in which $G$ is derived from an effective single-particle Hamiltonian constructed self-consistently using $\Sigma$ as the exchange-correlation contribution, which makes the result independent of the initial input. Although one advantage of the self-consistent GW is to eliminate the dependence of final results on the one-particle starting point, a fully self-consistent GW approach without vertex corrections has certain theoretical problems and corresponding calculations overestimate band gaps in semiconductors and insulators, and bandwidths in metals~\cite{Kutepov09,Kutepov17}. However, in many cases the so-called ``{\it one-shot}'' $\gowo$ method is deemed sufficiently accurate~\cite{Hybertsen85}, specially for the calculations of the band gap of solid hydrogen~\cite{McMinis}. Applying the first order vertex correction to the polarisibility and to the self-energy indicates that the vertex corrections and the diagrams dressing $G$ and $W$ cancel out to a high degree that provides a justification for the one-shot $\gowo$ approach~\cite{Engel}. Here, for Si and Ge, $G$ and $W$ are built with the exchange-correlation potential in the local-density approximation (LDA) of the density functional theory (DFT), and the self-energy determines the correction to the LDA quasiparticle energies, as explained in the next section. In the case of solid hydrogen, however, the main results were obtained with Becke-Lee-Yang-Parr (BLYP) approximation for exchange-correlation potential of DFT part, as it provides a better starting point for the many-body expansion due to having a smaller exchange-correlation self-interaction errors\cite{PRB13}.

Applying the finite temperature Green's function formalism to real systems with the "chemical accuracy" is a formidable task because the eigenvalue spectrum of realistic Hamiltonians is wide. In this work, we develop an efficient implementation of the GW space-time (GWST) method~\cite{Rojas, White, Rieger, Steinbeck} in which the functions $G$ and $W$ are constructed in real space and imaginary time at finite temperature. The quasi-particle excitation energies are typically obtained with the help of analytic continuation of the self-energy to the real frequency domain, which is a difficult problem in itself and a source of additional systematic errors. The finite-temperature formalism allows us to extract band gaps in a controlled way from the asymptotic behavior of the full many-body Green's function at large imaginary times. More generally, we demonstrate control of all systematic errors apart from that of the GW approximation, concluding that the employed Chebyshev polynomial representation~\cite{Gull18,JiaLi2020} suites this purpose quite well. An important feature of our approach is that the $G$ and $W$ functions are represented and stored in memory in full, which makes it amenable to immediate extension to higher orders of the diagrammatic expansion by means of stochastic sampling of the series using the diagrammatic Monte Carlo (DiagMC) technique~\cite{Prokofiev98, Kozik, Houcke, Mishchenko, van2012feynman,deng2015emergent,Chen19}. 

We apply our approach to evaluate the direct and indirect band gaps of the $hcp$ solid hydrogen under compression. To this end, we perform one-shot $\gowo$ calculations, the accuracy of which is expected to be comparable to that of more advanced GW schemes due to the chemically primitive composition of the system, involving only $s$ electrons.  We first validate this approach by reproducing benchmarks for the Si and Ge crystals, which are well understood both experimentally and numerically. We then study the energy band gap reduction in $hcp$ solid hydrogen as a function of pressure, and repeat the calculations using the fully self-consistent GW, which provides an indication of the systematic errors of the approximation. With a speculation on the role of lattice dynamics and vibrons in decreasing the band gap, we conclude that the solid hydrogen above 270 GPa cannot adopt the $hcp$ crystal structure.

The paper is organized as follows. Section~\ref{CD} consist of two parts. Part~\ref{GW} describes at the conceptual level the GWST method and our approach to the energy band gap calculation. In part~\ref{ND}, we discuss more technical details, including those of the underlying DFT calculations and numerical representation of the $G$ and $W$ functions in imaginary time. In Sec.~\ref{sec:newSchemes} we demonstrate our workflow for the band gap extraction from the Green's function asymptotic behaviour (Sec.~\ref{newQPengyEstimation}) and extrapolation to the infinity of the discretisation parameters (Sec.~\ref{sec:ExtrapolationRes}) using a test set of materials. We present and discuss our results for solid hydrogen in the context of available experimental data in Sec~\ref{results-H}. The section~\ref{con} concludes the paper. 

\section{\label{CD} The GW space-time method}
\subsection{\label{GW} General formulation}
In the GWST method~\cite{Rojas, White, Rieger, Steinbeck}, the Green's function ($G$), polarizability ($P$), dielectric function ($\varepsilon$), dynamically screened Coulomb interaction ($W$), and self-energy ($\Sigma$) are defined and stored in the real space (parameterized by the radius-vector $\mathbf{r}$) and imaginary time ($\tau$) at a given temperature $T$, typically on appropriately constructed grids. The first approximation $G_0(\boldsymbol{rr}'\tau)$ to the full interacting Green's function is constructed from the effective Hamiltonian where the electron-electron interactions are described by the exchange-correlation potential of the DFT~\cite{Hybertsen85} with the help of a standard approximation, such as, e.g., the LDA. 

We define
\begin{equation}
G_0(\boldsymbol{rr}'\tau)=\sum_{k} G_{k}^{0}(\tau)\psi_{k}(\boldsymbol{r})\psi_{k}^{*}(\boldsymbol{r}') \label{eq:G0_rr_KS}
\end{equation}
with $G_{k}^{0}(\tau)$ the Green's function in the Kohn-Sham (KS) basis, which has the diagonal form
\begin{eqnarray}
G_{k}^{0}(\tau)
&=&
    -\theta( \xi_{k})(1-f_{\beta}(\xi_{k}))e^{-\xi_{k}\tau}\nonumber\\
&&
    -\theta(-\xi_{k})   f_{\beta}(\xi_{k}) e^{ \xi_{k}(\beta-\tau)}, \label{eq:G0_ks}
\end{eqnarray}
where $k$ stands for the combined band- ($l$) quasimomentum ($\mathbf{k}$) index $k=\{l,\boldsymbol{k}\}$, $f$ is the Fermi distribution function, and $\xi_{k}$ and $\psi_{k}(\boldsymbol{r})$ are the KS Hamiltonian eigenvalues and eigenfunctions respectively. 

The irreducible polarizability is then found within the random phase approximation,
\begin{equation}
P(\boldsymbol{rr}'\tau)=G_0(\boldsymbol{rr}'\tau)G_0^{*}(\boldsymbol{rr}',-\tau)\label{eq:polar}
\end{equation}
and Fourier transformed to the reciprocal space, $P(\boldsymbol{rr}'\tau)\rightarrow P(\boldsymbol{qGG}'\tau)$, where $\boldsymbol{q}$ and $\boldsymbol{G}$, $\boldsymbol{G}'$ are the quasimomentum in the Brillouin zone and reciprocal lattice vectors correspondingly. To determine the screened interaction $W$ from the polarizability via the Dyson equation algebraically, the function $P(\boldsymbol{qGG}'\tau)$ needs to be converted to the domain of the Matsubara frequency $i\omega$ by a Fourier transform (FT) $P(\boldsymbol{qGG}'\tau) \rightarrow P(\boldsymbol{qGG}'i\omega)$. The FT implementation depends on how the dependence of the imaginary time is represented: when a variable-step grid is used~\cite{Ku}, the FT typically requires an interpolation onto a more suitable for the FT grid; in the Chebyshev-polynomial representation~\cite{Gull18,JiaLi2020} this extra operation is not needed for the FT. The specific details are discussed in the following section. 

The inverse dielectric function, which accounts for dynamical screening of the Coulomb interaction, straightforwardly follows from $P(\boldsymbol{qGG}'i\omega)$ with the symmetrized version given as
\begin{equation}
\varepsilon^{-1}(\boldsymbol{qGG}',i\omega)=\left[\delta_{\boldsymbol{GG}'}+4\pi
\frac{P(\boldsymbol{qGG}',i\omega)}{|\boldsymbol{q}+
\boldsymbol{G}||\boldsymbol{q}+\boldsymbol{G}'|}\right]^{-1},
\end{equation} 
and after the inverse FT $\varepsilon^{-1}(\boldsymbol{qGG}',i\omega)\rightarrow\varepsilon^{-1}(\boldsymbol{qGG}',\tau)$ yields
\begin{equation}
W(\boldsymbol{qGG}'\tau)=\sum_{\boldsymbol{qGG}'}\frac{4\pi}{\Omega}\frac{\varepsilon^{-1}(\boldsymbol{qGG}',\tau)}{|\boldsymbol{q}+\boldsymbol{G}||\boldsymbol{q}+\boldsymbol{G}'|},
\end{equation}
where $\Omega$ is the volume of the unit cell. Finally, the screened interaction is transformed to the direct space $W(\boldsymbol{qGG}'\tau)\rightarrow W(\boldsymbol{rr}'\tau)$, so that the self-energy operator, $\Sigma(\boldsymbol{rr}'\tau)=-G(\boldsymbol{rr}'\tau)W(\boldsymbol{rr}'\tau)$, and its matrix elements 
$\Sigma_{lm\kk}(\tau)=\left\langle \psi_{l\kk}|\Sigma(\boldsymbol{rr}'\tau)|\psi_{m\kk}\right\rangle $ could be computed. For further analysis, the self-energy matrix, denoted by $\hat{\Sigma}_{\kk} \equiv \{\Sigma_{lm\kk}\}$, in converted to the Matsubara frequency space by yet another FT, $\hat{\Sigma}_{\kk}(\tau)\rightarrow\hat{\Sigma}_{\kk}(i\omega)$.

We use two methods to extract the quasi-particle excitation energies. The first, traditional one, relies on analytic continuation of the self-energy to real frequencies. It is an intrinsically approximate approach, the accuracy of which is generally difficult to control since numeric analytic continuation is fundamentally an ill-defined problem. However, numerically stable and physically meaningful results have been obtained by first neglecting the off-diagonal matrix elements, which is typically justified for semiconductors, and finding the diagonal at real frequencies $\Sigma_{ll\kk}(\omega)$ via fitting the $\Sigma_{ll\kk}(i\omega)$ to a multipole expansion model~\cite{Rojas,White,Rieger}. This constitutes the simplest analytic continuation model. More general, Pad{\'{e}} approximant technique suffers the problems of numerical stability. It is worth to mention other analytic continuation methods, more stable than the plain Pad{\'{e}}, such as the averaged Pad{\'{e}} approximant technique~\cite{Schott2016} and the Thiele’s reciprocal difference method applied in Ref.~\cite{Liu2016}. The quasi-particle (QP) energies are then found as solutions of the equation:
\begin{eqnarray}
\xi_{l\kk}^{qp}=\xi_{l\kk}+\text{Re}[\Sigma_{ll\kk}(\omega & = & \xi_{l\kk}^{qp})]-v_{l\kk}^{xc}
\label{eq:xi_qp}
\end{eqnarray}
where $v_{l\kk}^{xc}$ stands for the exchange-correlation potential used in the effective Hamiltonian to construct $G_0$, which needs to be subtracted to avoid double counting.

\begin{figure}
\includegraphics[width=0.48\textwidth]{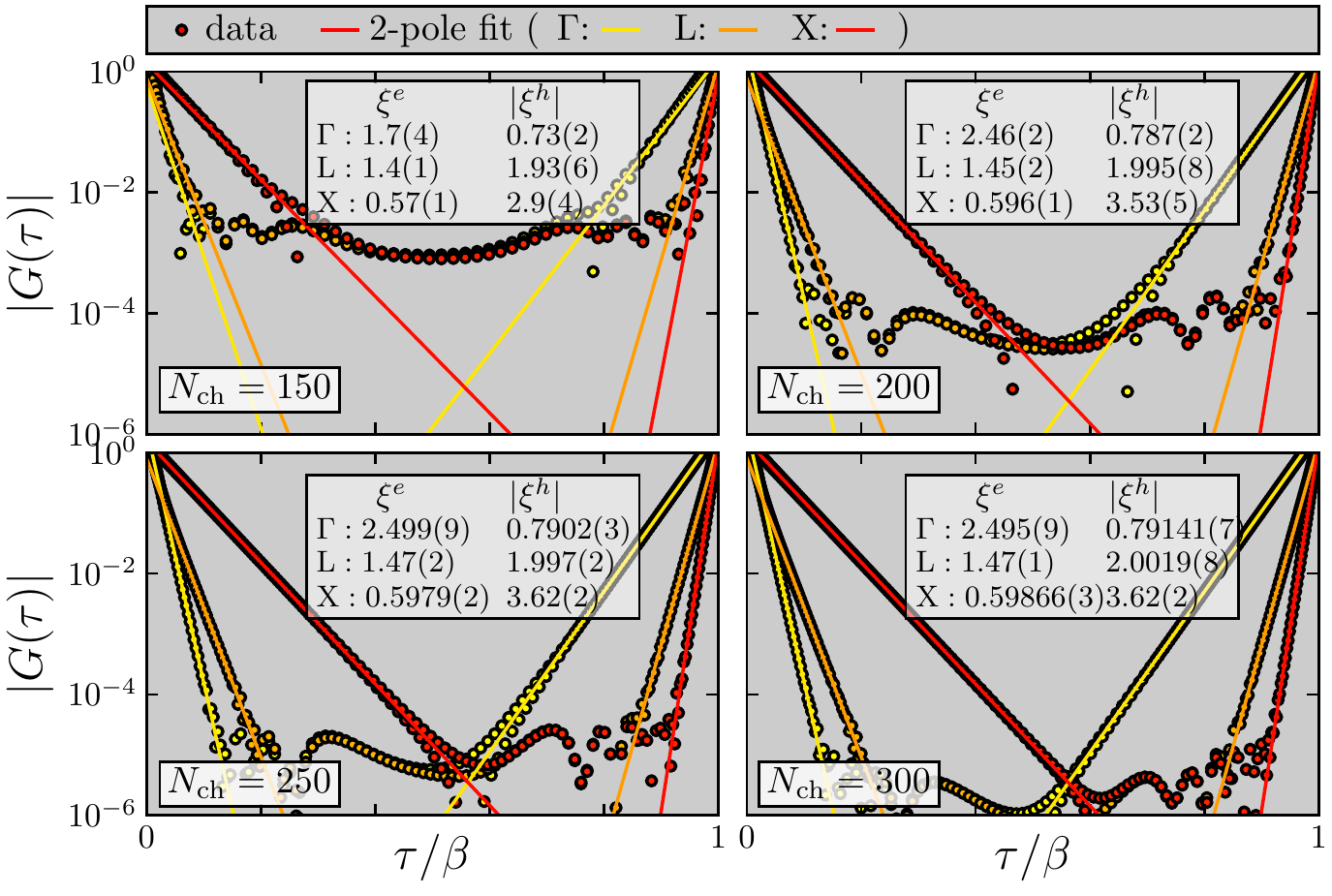}
\caption{\label{cheb} The dependence of $G_{\kk}$ on the imaginary time $\tau$ at high-symmetry k-points, and the fit of its asymptotic behavior, Eq.~(\ref{eq:giik_traced}), used to extract the  valence- ($\xi^{h}$) and conductance- ($\xi^{e}$) band energies for Si shown as unframed numbers (in eV) with error bars in brackets. Note, that left(right) columns of quasiparticle energies correspond to left(right) slopes of $G_{\bf k}(\tau)$ curves. Different panels correspond to different numbers of Chebyshev polynomials (framed numbers) used to represent $\hat{G}_{\kk}(\tau)$. Color of a fitted line identifies a k-point in the way shown in the upper panel. Temperature was set to $300$ K.}
\end{figure}

This approach is useful for providing an estimate of the QP spectrum, but, for the problem of determining the band gap, the additional approximation introduced by the analytic continuation is not necessary. The energy of excitations nearest to the Fermi level can be found in a controlled way from the asymptotic decay of the full Green’s function at long imaginary times:
\begin{eqnarray}
|G_{\kk}(\tau)|\rightarrow
    a^{e}_{\kk} e^{-\xi^{e}_{\kk}\tau}
   +a^{h}_{\kk} e^{ -|\xi^{h}_{\kk}|(\beta-\tau)},
   \label{eq:giik_traced}\\
   \tau \gg [\xi^{e}]^{-1},\ \ \ [\beta - \tau] \gg |\xi^{h}|^{-1}\nonumber
\end{eqnarray}
where $a^{e(h)}_{\kk}$ are some constants and $\xi^{e(h)}$ are the energies of the electron(hole)-like excitations. Here $G_{\kk}$ can be taken as a sum over band indices of the full Green's function matrix,
\begin{equation}
    G_{\kk}(\tau)=\sum_{lm} G_{lm\kk}(\tau), \label{eq:G_k}
\end{equation}
where $G_{lm\kk}(\tau)$ is obtained by solving the Dyson equation
\begin{equation}
    \hat{G}_{\kk}(i\omega)=\left[i\omega\hat{1}-(\hat{\Sigma}_{\kk}(i\omega)-\hat{v}_{\kk}^{xc})\right]^{-1}\label{eq:Dyson}
\end{equation}
with the subsequent FT $\hat{G}_{\kk}(i\omega)\rightarrow\hat{G}_{\kk}(\tau)$, where the matrix notation $\hat{G}_{\kk} \equiv \{G_{lm\kk}\}$ is used and $\hat{1}$ is the unity matrix. Following Eq.~(\ref{eq:giik_traced}), the valence- ($\xi^{h}$) and conductance- ($\xi^{e}$) band energies  can be extracted as linear fits of $\log[|G(\tau)|]$, as exemplified in Fig.~\ref{cheb} (discussed in more detail below). 

Although the proposed scheme allows to extract the lowest excitation energies, other energies can, in principle, be computed, first, by fitting the untraced $G_{lm\kk}(\tau)$ at diagonal. This corresponds to the assumption that the Kohn-Sham orbitals are a good approximation to the true quasiparticle ones. In that cases, where such approximation breaks down, one would have to apply a transformation of the Green's function to the basis of quasiparticle orbitals estimated, for example, via the full quasiparticle equation with an analytically continued self-energy. It is not clear, however, if such a scheme should give an improvement over the pure analytic continuation one, but it is worth the future studying.

The results obtained from Eq.(\ref{eq:xi_qp}) with the analytic continuation of the self-energy and those found from the asymptotic form of the Green's function (\ref{eq:giik_traced}) will be referred to as $\gac$ and $\ggf$, respectively. We compare the energy band gaps extracted by both methods in the benchmark study of Si, Ge and H in Sec.~\ref{sec:ExtrapolationRes}. 

\subsection{Details of implementation}\label{ND}
The building-block functions (G, P, W, and $\Sigma$) in the GWST technique decay fast with increasing the distance between ${\bf r}$ and ${\bf r}^\prime$, except for W$({\bf r}, {\bf r}^\prime)$, which is long-ranged with $\sim 1/|{\bf r}-{\bf r}^\prime|$ behavior. When storing functions of ${\bf r}$ and ${\bf r}^\prime$, following the original nomenclature~\cite{Rojas, Rieger}, we define the domain of the first argument ${\bf r}$ to be the unit cell (UC), which is put on a mesh of $N_r$ points. The second argument ${\bf r}^\prime$ is defined within the so-called interaction cell (IC). It is a larger domain with periodic boundary conditions, consisting of $N_k$ unit cells. The size of the interaction cell defines the corresponding quasimomentum mesh in the Brillouin zone of $N_k$ points. The specific crystal symmetries are used to reduce the memory consumption by storing all quantities on the irreducible wedge of ${\bf r}$ coordinate grid, while  ${\bf r}^\prime$ coordinate grid always has $N_r \times N_k$ size, spanning the IC. 

\textit{Chebyshev polynomial (CP) representation} provides an easily controllable way of storing and manipulating correlation functions in both time and frequency domains, as well as the Fourier transform algebra for switching between them. The concept is essentially that of representing a continuous function $F(\tau)$ as an expansion in an orthonormalized basis of polynomials $\{T_l(\tau)\}$, with the error controlled by the number of basis functions $N_{ch}$ used in the representation:
\begin{eqnarray}
  F(\tau)
  &=&\sum_{l=0}^{N_{ch}-1} F_lT_l(\tau),\label{eq:cheb_tau}\\
  F(i\omega)
  &=&\sum_{l=0}^{N_{ch}-1} F_lT_l(i\omega),\label{eq:cheb_iw}\\
  T_l(i\omega)
  &=&\int_{0}^{\beta} d\tau T_l(\tau) e^{i\omega\tau}. \label{eq:cheb_tauiw}
\end{eqnarray}
Once the coefficients $F_l$ are obtained, computing the Fourier transform is trivial with the help of the pre-tabulated functions $T_l(\tau)$ and $T_l(i\omega)$.

This framework has been discussed in detail in Refs.~\cite{Gull18,JiaLi2020} and applied to diverse systems in Ref.~\cite{JiaLi2020}. The methodology combines an accuracy and simplicity and, therefore, can be straightforwardly used for the total energy calculation~\cite{JiaLi2020}. One should note that similar ideas are behind the GW implementation in the series of works~\cite{Kutepov2009,Yin2011,Kutepov2012,Kutepov16,Kutepov17}, where the Chebyshev polynomials are used as well, with, however, more sophisticated workflow with more control parameters. Another advanced scheme is known, the compressive sensing approach discussed in Ref.~\cite{Kaltak2020}, which allows to greatly compress the representation of the correlation functions, and is worth the future implementation. 

It was shown in the original works~\cite{Gull18,JiaLi2020} that for the basis $\{T_l(\tau), l=0,\ldots, N_{ch}-1\}$ the optimal time mesh for sampling $F(\tau)$ in the calculation of $F_l$, must include exactly $N_{ch}$ points:
\begin{eqnarray}
  \bar{\tau}_n=\tau\left[\cos\left(\pi\frac{2n+1}{2N_{ch}}\right)\right], \ \ \ n=0 \ldots N_{ch}-1
\end{eqnarray}
which are the roots of the higher-order polynomial $T_{N_{ch}}(\tau)$. Here $\tau[x]$ function is a linear map from $[-1:1]$ to $[0:\beta]$: $\tau[x]=\beta(x+1)/2$, $-1 \leq x \leq 1$. Similarly, the points $i\bar{\omega}_n$, $n=0 \ldots N_{ch}-1$, for sampling the imaginary frequency dependence $F(i\omega)$ are chosen as roots of the Fourier transform counterpart $T_{N_{ch}}(i\omega)$, with a minor difference between bosonic and fermionic cases~\cite{JiaLi2020}. The polynomial coefficients $F_l$ are then obtained, either from the given imaginary-time ($F(\tau)$) or frequency ($F(i\omega)$) dependence, according to
\begin{eqnarray}
    F_l &=&\sum_{n=0}^{N_{ch}-1} [A^{-1}]_{ln} F(\bar{\tau}_n)\nonumber \\
        &=&\sum_{n=0}^{N_{ch}-1} [\widetilde{A}^{-1}]_{ln}F(i\bar{\omega}_n),
        \label{eq:F_l_from_tauomega}
\end{eqnarray}
with the transformation matrices $A$ and $\widetilde{A}$ of size $N_{ch}\times N_{ch}$ defined by
\begin{eqnarray}
   A_{ln}&=&T_{l}(\bar{\tau}_n), \nonumber\\
   \widetilde{A}_{ln}&=&T_{l}(i\bar{\omega}_n).\label{eq:Tl_iwn}
\end{eqnarray}
The coefficients $F_l$ are sufficient to obtain both $F(\tau)$ or $F(i\omega)$ at any given argument through Eqs.~(\ref{eq:cheb_tau}), (\ref{eq:cheb_iw}).

\begin{figure}
\includegraphics[width=0.48\textwidth]{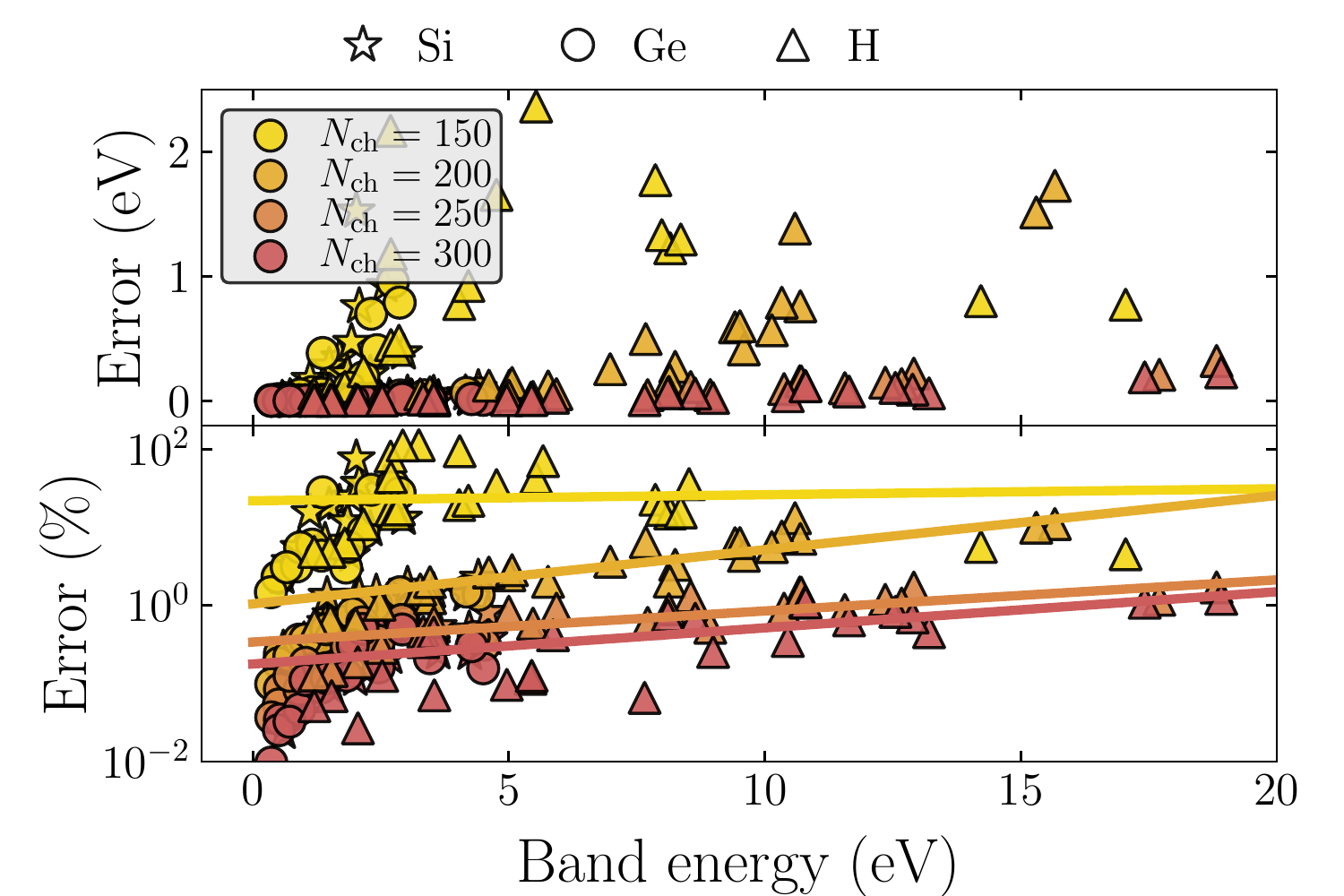}
\caption{\label{absAndRelError} Collection of (absolute)energy/error data points obtained by the fitting procedure depicted in Fig.~\ref{cheb} repeated for Si, Ge and H across the BZ momentum points. Both absolute (top) and relative (bottom) errors are plotted as functions of band energy. Solid lines represent the fits to $f(x)= ae^{bx}$ functions of band energy $x$.}
\end{figure}

The obvious disadvantage of this framework is that resolving the time dependence due to the states far away from the Fermi level requires a larger number of polynomials and corresponding grid points $\{\bar{\tau}_n\}$. This means that capturing the effects of core states or extremely high-energy empty states on the imaginary time dependence could be memory demanding. Having 100 bands for Si, we have to use 250 polynomials to achieve negligible discretisation errors. However, the advantage of the CP framework is that, once the result is converged with $N_{ch}$, the systematic error of the FT is automatically negligible. Secondly, all time/frequency dependent functions are well represented in all regions of time/frequency axis uniformly, which is essential for obtaining energies from the asymptotic behavior, Eq.~\ref{eq:giik_traced}. In addition, the Fourier transform $G(i\omega)\rightarrow G(\tau)$ generally requires a careful treatment of the high-frequency tail due to the slow convergence of the truncation error, but in the CP representation the correct asymptotic is build in automatically. 
 
Although the temperature was set to $300$ K in all Matsubara time/frequency grids throughout this work, the presented framework can be used in a wide temperature range. The numerical cost for construction of Chebyshev polynomials in the imaginary-frequency domain grows with decreasing temperature due to a larger size of unsparsed Matsubara frequency grid for finding the roots of the  $N_{ch}$'th polynomial $T_{N_{ch}}(i\omega)$. This, however, is performed once before the GW calculations themselves, and therefore, the framework can be used even in a sub-Kelvin regime.

Finally, the scheme may find its applications, for example, in the diagrammatic Monte-Carlo~\cite{Prokofiev98, Kozik, Houcke, Mishchenko, van2012feynman,deng2015emergent,Chen19} calculations, where the correlation functions contain a numerical noise, and AC-based approaches become unstable.

\section{Results\label{sec:newSchemes}}
\subsection{\label{newQPengyEstimation} Controlled estimation of QP energies}
We demonstrate a procedure allowing for extraction of the lowest QP energies from the imaginary-time GF asymptotic. We use Si and Ge as a validation set, and compare the resulting bandgap values with available experimental data, but also consider the solid hydrogen under pressure (discussed in details in Sec.~\ref{results-H}) for determination of the error bars. Figure~\ref{cheb} shows an example of the imaginary-time dependence of $G_\mathbf{k}$, Eq.~(\ref{eq:G_k}), for Si computed with different values of $N_{ch}$ and the result of the fit to the asymptotic form (\ref{eq:giik_traced}) used to determine the band gap in the $\ggf$ approach. The fitting error behaves smoothly with the number of polynomials $N_{ch}$ used in the CP representation. This can be seen when repeating the procedure shown in Fig.\ref{cheb} for fitting the set of lowest QP energies for all studied materials over the Brillouin Zone k-points. We see (Fig.~\ref{absAndRelError}) that the relative error has about two orders of magnitude variation with the band energy, not exceeding, however $0.9\%$ $0.7\%$ for Si and Ge respectively, and $0.3\%$ for energy states within the relevant $[-3:3]$ eV window for $94$ GPa pressurised hydrogen when taking $N_{ch}=250$. It is evident that the error bar values reduce with decreasing the absolute of associated band energy values, and estimation at $94$ GPa in H case will give the upper error limit at higher pressures due to the reduction of the solid hydrogen's energy gap with pressure (e.g., see Sec.~\ref{results-H}). Together with the extrapolation error discussed in the next section, these relative error estimations will be used when reporting the final QP energies and gaps, and consequently, all subsequent calculations will be done with $N_{ch}=250$. 

\subsection{Extrapolation to infinite grid sizes\label{sec:ExtrapolationRes}}
\begin{figure}
    \includegraphics[width=0.49\textwidth]{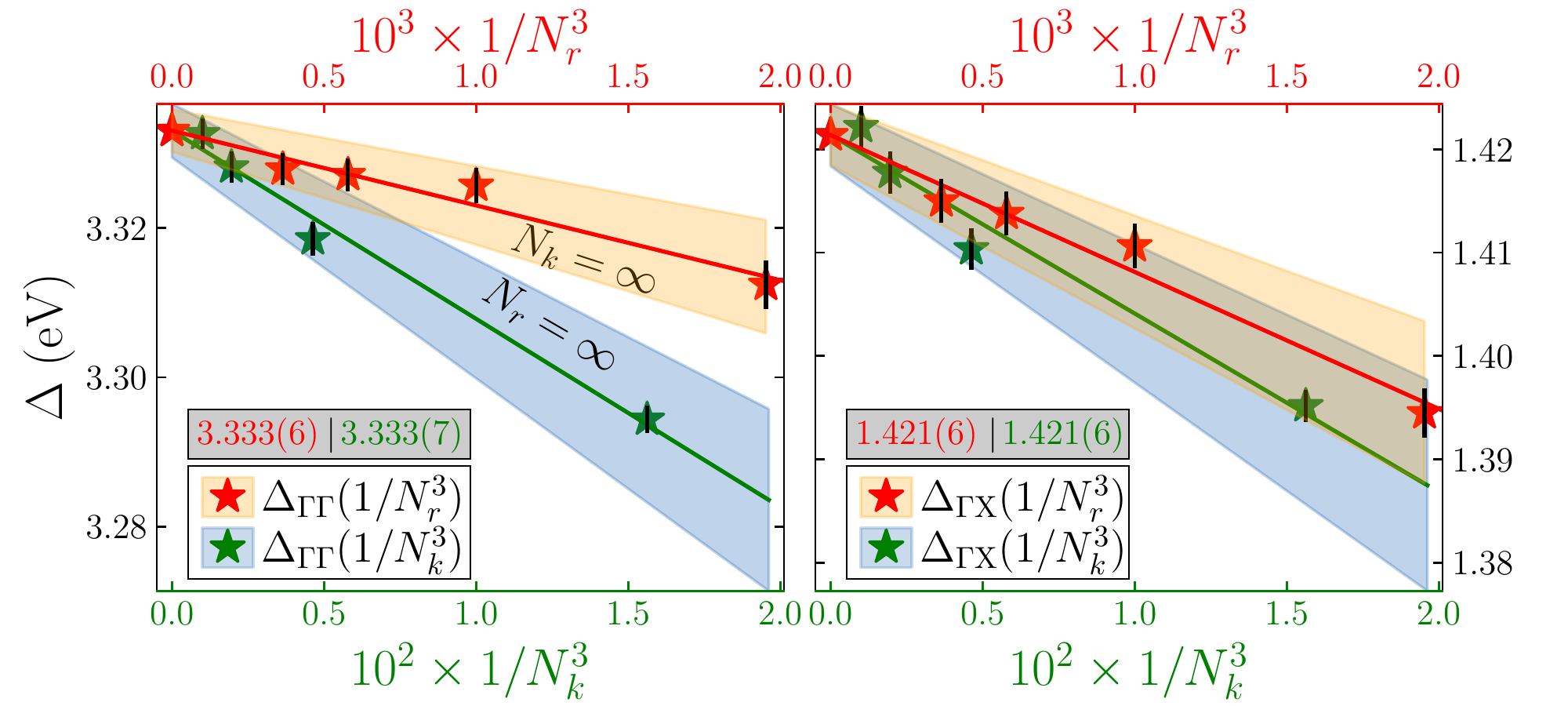}
    \caption{\label{Si_FS} 
        Illustration of the algorithm behind the final extrapolation of $\Gamma \rightarrow \Gamma$ (left) and $\Gamma \rightarrow X$ (right) bandgaps of Si to the infinite $N_r$ and $N_k$ grid parameters. Green/red data points correspond to $N_{r/k}=\infty$ columns/rows of the Tab.~\ref{gwgap_CH} ($\ggf$ data) both carrying the data extrapolation error bars. The linear fit of these values to a $\sim\frac{1}{N_{k/r}}$ function is given by lines with associated fitting error bars shown by the shaded regions. The size of the shaded region is chosen to cover both the error of the fit itself and all error bars of data points.
}
\end{figure}

\begin{figure}
    \includegraphics[width=0.49\textwidth]{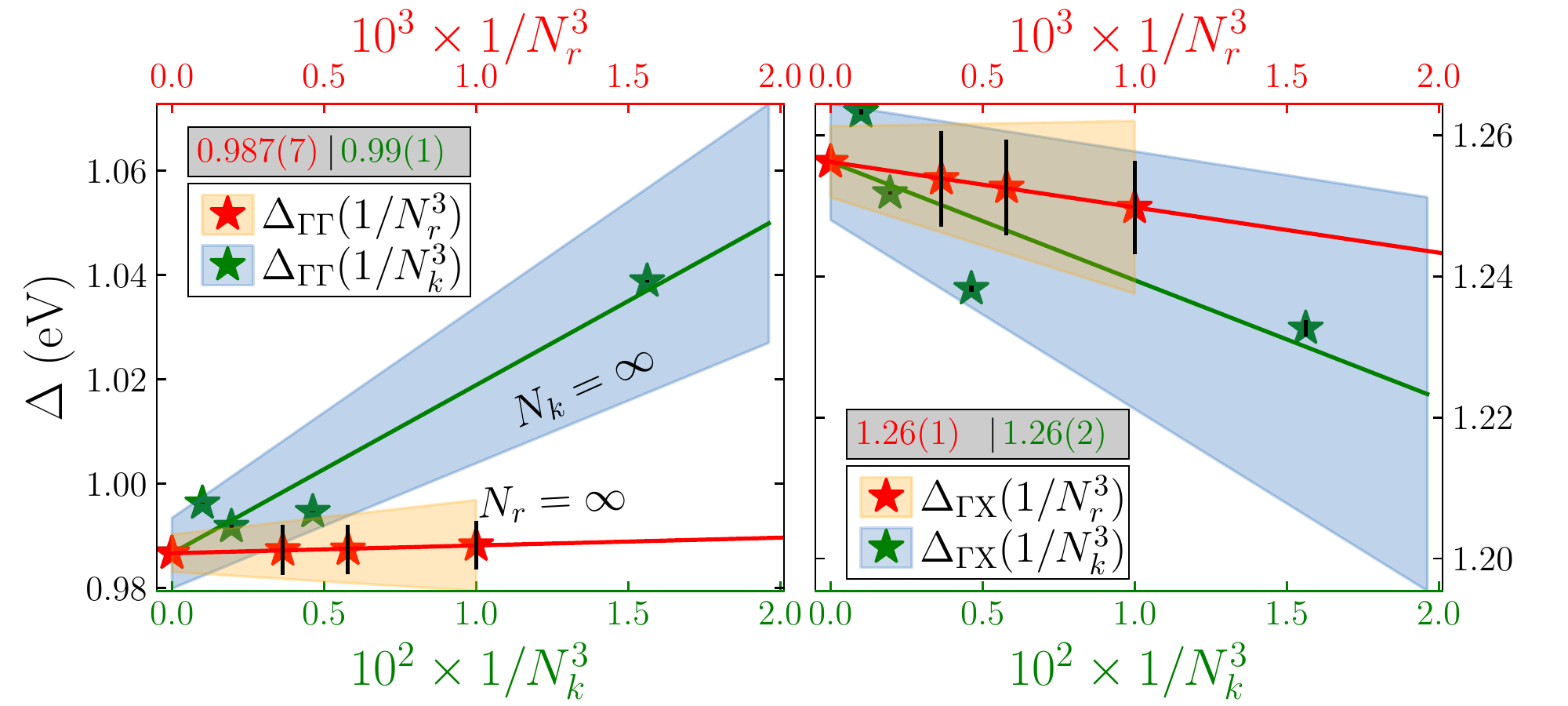}
    \caption{\label{Ge_FS} 
        Same as figure~\ref{Si_FS} but for Ge.
    }
\end{figure}

\begin{figure}
    \includegraphics[width=0.49\textwidth]{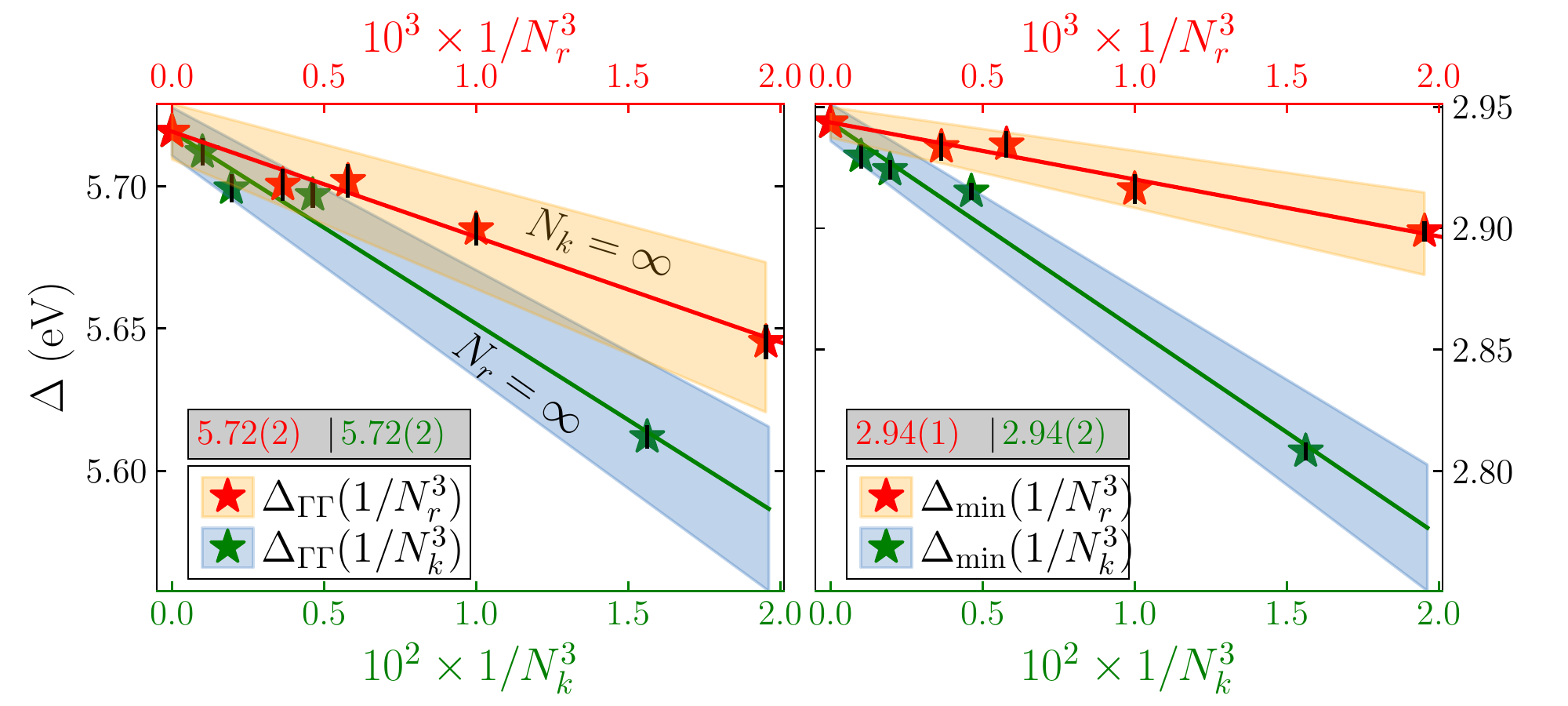}
    \caption{\label{H_FS} 
        Same as figure~\ref{Si_FS} but for H with, however, a slightly different scheme for estimation of the $\Delta_\mathrm{min}$ (see Sec.~\ref{sec:ExtrapolationRes})
    }
\end{figure}

\begin{table}
 \begin{tabular}{|c|c|c|c|c|c|}
 \hline
 k-mesh & $N_r=8^3$& $N_r=10^3$& $N_r=12^3$ & $N_r=14^3$  & $N_r=\infty$  \\
 \hline\hline
 \multicolumn{6}{|c|}{$\Gamma \rightarrow X$: Si-$\gac$} \\
 \hline
  $N_k=4^3$ & 1.325 & 1.341 & 1.344 & 1.346 & 1.352(4)\\
  $N_k=6^3$ & 1.336 & 1.352 & 1.356 & 1.357 & 1.363(4)\\
  $N_k=8^3$ & 1.342 & 1.357 & 1.361 & 1.362 & 1.368(4)\\
  $N_k=10^3$ & 1.345 & 1.361 & 1.364 & 1.365 & 1.371(4)\\
  $N_k=\infty$ & 1.344(3) & 1.360(3) & 1.363(2) & 1.364(2) & 1.371(6) \\

 \hline
 \multicolumn{6}{|c|}{$\Gamma \rightarrow X$: Si-$\ggf$} \\
 \hline
  $N_k=4^3$ & 1.372 & 1.385 & 1.389 & 1.390 & 1.395(3)\\
  $N_k=6^3$ & 1.384 & 1.400 & 1.403 & 1.404 & 1.410(4)\\
  $N_k=8^3$ & 1.391 & 1.407 & 1.410 & 1.411 & 1.418(4)\\
  $N_k=10^3$ & 1.396 & 1.412 & 1.415 & 1.416 & 1.422(4)\\
  $N_k=\infty$ & 1.394(5) & 1.411(4) & 1.414(4) & 1.415(4) & 1.421(6)\\
 \hline
 \multicolumn{6}{|c|}{$\Gamma \rightarrow X$: Ge-$\gac$} \\
 \hline
  $N_k=4^3$ && 1.192 & 1.195 & 1.196 & 1.200(1)\\
  $N_k=6^3$ && 1.218 & 1.222 & 1.223 & 1.2257(9)\\
  $N_k=8^3$ && 1.228 & 1.232 & 1.233 & 1.2359(9)\\
  $N_k=10^3$ && 1.235 & 1.238 & 1.239 & 1.2423(8)\\
  $N_k=\infty$ && 1.235(5) & 1.238(5) & 1.239(5) & 1.242(7) \\
  \hline
  \multicolumn{6}{|c|}{$\Gamma \rightarrow X$: Ge-$\ggf$} \\
  \hline
  $N_k=4^3$ && 1.223 & 1.228 & 1.229 & 1.233(2)\\
  $N_k=6^3$ && 1.231 & 1.235 & 1.236 & 1.2382(8)\\
  $N_k=8^3$ && 1.245 & 1.248 & 1.249 & 1.2518(4)\\
  $N_k=10^3$ && 1.256 & 1.260 & 1.261 & 1.2633(8)\\
  $N_k=\infty$ && 1.25(1) & 1.25(1) & 1.25(1) & 1.26(2) \\
  \hline
  \multicolumn{6}{|c|}{min. gap: H-$\gac$ (94 GPa)} \\
  \hline
  $N_k=4^3$ & 2.751 & 2.765 & 2.781 & 2.789 & 2.79(1)\\
  $N_k=6^3$ & 2.849 & 2.868 & 2.885 & 2.884 & 2.895(7)\\
  $N_k=8^3$ & 2.851 & 2.868 & 2.886 & 2.884 & 2.894(8)\\
  $N_k=10^3$ & 2.842 & 2.853 & 2.872 & 2.871 & 2.88(1)\\
  $N_k=\infty$ & 2.86(2) & 2.88(3) & 2.90(3) & 2.89(3) & 2.90(4) \\

  \hline
  \multicolumn{6}{|c|}{min. gap: H-$\ggf$ (94 GPa)} \\
  \hline
  $N_k=4^3$ & 2.770 & 2.785 & 2.801 & 2.799 & 2.808(7)\\
  $N_k=6^3$ & 2.868 & 2.889 & 2.905 & 2.904 & 2.915(7)\\
  $N_k=8^3$ & 2.878 & 2.897 & 2.915 & 2.913 & 2.924(8)\\
  $N_k=10^3$ & 2.890 & 2.903 & 2.922 & 2.921 & 2.93(1)\\
  $N_k=\infty$ & 2.899(8) & 2.92(1) & 2.93(1) & 2.93(1) & 2.94(2) \\
  \hline
\end{tabular}
\caption{\label{gwgap_CH} $\Gamma \rightarrow X$ energy band gap of Si and Ge (in eV) and minimal gap of pressurised hydrogen (more details in Sec.~\ref{results-H}) obtained by $\gac$ and $\ggf$ techniques at $T = 300 K$ and at different quasimomentum and real space grids. The $N_{k/r}=\infty$ rows/columns are extrapolations of columns/rows of data to the infinite grids using the linear to $\frac{1}{N_{r/k}}$ fitting function. The final value for the gap, which is written into the cell of $N_{k/r}=\infty$ row/column crossing, is taken as an additional extrapolation of either $N_k=\infty$ row or $N_r=\infty$ column, the one with the largest error bar. The algorithm behind the final error bar estimation is visually illustrated in Figs.~\ref{Si_FS}-\ref{H_FS}. Error bar of the time/frequency representation is not taken into account explicitly in this table.
}
\end{table}
Table.~\ref{gwgap_CH} shows that the AC and GF bandgaps may differ by a $\sim50$ meV (Tab.~\ref{gwgap_CH}), which is due to both GF fitting and AC error bars (note, that it is not possible to assess the AC error bar, as was discussed in Sec.~\ref{GW}). Figs.~\ref{Si_FS}-\ref{H_FS} illustrate the convergence patterns of the $N_{k/r}=\infty$ rows/columns of Tab.~\ref{gwgap_CH} and the methodology behind the extrapolation to the infinity of both parameters for all studied compounds.

The estimation of the minimal gap of H is done by first interpolating the extracted QP energies from a regular k-point grid to the fine-sampled path connecting the high-symmetry reciprocal k-points and then by searching for the minimal gap value on the path. It is hard to assess the interpolation error bar explicitly. However, due to the fact that our extrapolation error bars are the same for direct (where no interpolation needed) and minimal gaps for pressurised hydrogen, we can safely ignore this contribution.

Finally, our extrapolated gap values are compared with known results for Si and Ge in Tab.~\ref{compare}.
\begin{table}
 \begin{tabular}{|c|c|c|c|}
 \hline\hline
 \multicolumn{3}{|c|}{Si} \\
 \hline
 Method & $\Gamma \rightarrow \Gamma$ & $\Gamma \rightarrow X$ \\
 \hline
 space-time, GF (this work)  & 3.33(4) & 1.42(2)\\
 space-time 1 (Ref.~\onlinecite{Rieger})& 3.24 & 1.34\\ 
 space-time 2 (Ref.~\onlinecite{Rieger})& 3.32 & 1.42\\
 plane-wave (Ref.~\onlinecite{Hybertsen}) & 3.35& 1.44 \\
 plane-wave (Ref.~\onlinecite{Gruning}) &3.2 &  1.2 \\
 all-electron (Ref.~\onlinecite{Ku}) & 3.12  & \\
 all-electron (Ref.~\onlinecite{Hamada}) & 3.30  & \\
 all-electron (Ref.~\onlinecite{Arnaud}) & 3.15  & \\
 all-electron (Ref.~\onlinecite{Markprb06})& 3.16  & 1.11 \\
 Exp. (Ref.~\onlinecite{Hellwege})& 3.40  & \\
 Exp. (Ref.~\onlinecite{Ortega})& 3.05  & \\
 Exp. (Ref.~\onlinecite{Hybertsen}) & 3.4 & 1.3  \\
 Exp. (Ref.~\onlinecite{Gruning}) & 3.37 & 1.25\\
 \hline
 \hline\hline
 \multicolumn{3}{|c|}{Ge} \\
 \hline
 Method & $\Gamma \rightarrow \Gamma$ & $\Gamma \rightarrow X$ \\
\hline 
 space-time, GF (this work) & 0.99(2) & 1.26(3) \\
 plane-wave (Ref.~\onlinecite{Shirley})& & 1.23\\
 all-electron (Ref.~\onlinecite{Ku}) &1.11 &0.49 \\
 all-electron (Ref.~\onlinecite{Markprb06})& 0.84 & 0.97 \\
 Exp. (Ref.~\onlinecite{Zahlenwerte}) &1.0 &1.3 \\ 
 Exp. (Ref.~\onlinecite{Shirley}) &0.89& 1.10\\
 \hline\hline
\end{tabular}
\caption{\label{compare} A comparison between the $\Gamma \rightarrow \Gamma$ and $\Gamma \rightarrow X$ $\ggf$ band gaps of Si and Ge together with a slice of theoretical $\gowo$ and experimental results. Energies are in eV. Error bars associated with bandgaps are composed out of extrapolation to infinite grids and $\ggf$ band energy extraction error bars.}
\end{table}

\section{\label{results-H} Results: Solid Hydrogen}
We considered solid molecular structure with hexagonal closed pack $hcp$ symmetry $(P6_3/mmc)$ with four hydrogen atoms per primitive cell. Our DFT calculations were carried out using the Quantum-Espresso (QE) suite of programs~\cite{QE} and  the Becke-Lee-Yang-Parr (BLYP)~\cite{BLYP} XC-DFT functional. We used a basis set of plane waves with an energy cutoff of 100~Ry. For the geometry and cell optimisations a $16\times16\times16$ ${\bf k}$-point mesh with the QE-tabulated standard norm-conserving pseudopotential is used. The quasi-Newton algorithm was used for all cell and geometry optimization, with a convergence thresholds on the total energy and forces of 0.01 mRy and 0.1 mRy/Bohr, respectively. The DFT-BLYP band structure is calculated using a dense k-mesh of $24\times24\times24$.  

Recent X-ray diffraction experiment~\cite{CJi} of high-pressure hydrogen, observed three Bragg peaks which are consistent with an $P6_3/mmc$ structure with $c/a$ ratio close to $\sqrt{8/3}$, and the studied structure was reported as an isostructural $hcp$. According to their electronic band structure study the conduction band minimum occurs at the $\Gamma$ point and the valence band maximum was found at the $A$ and $K$ high-symmetry k-points with indirect band gap of 3.8 eV at 100 GPa.

\begin{figure}
\centering
\includegraphics[width=0.49\textwidth]{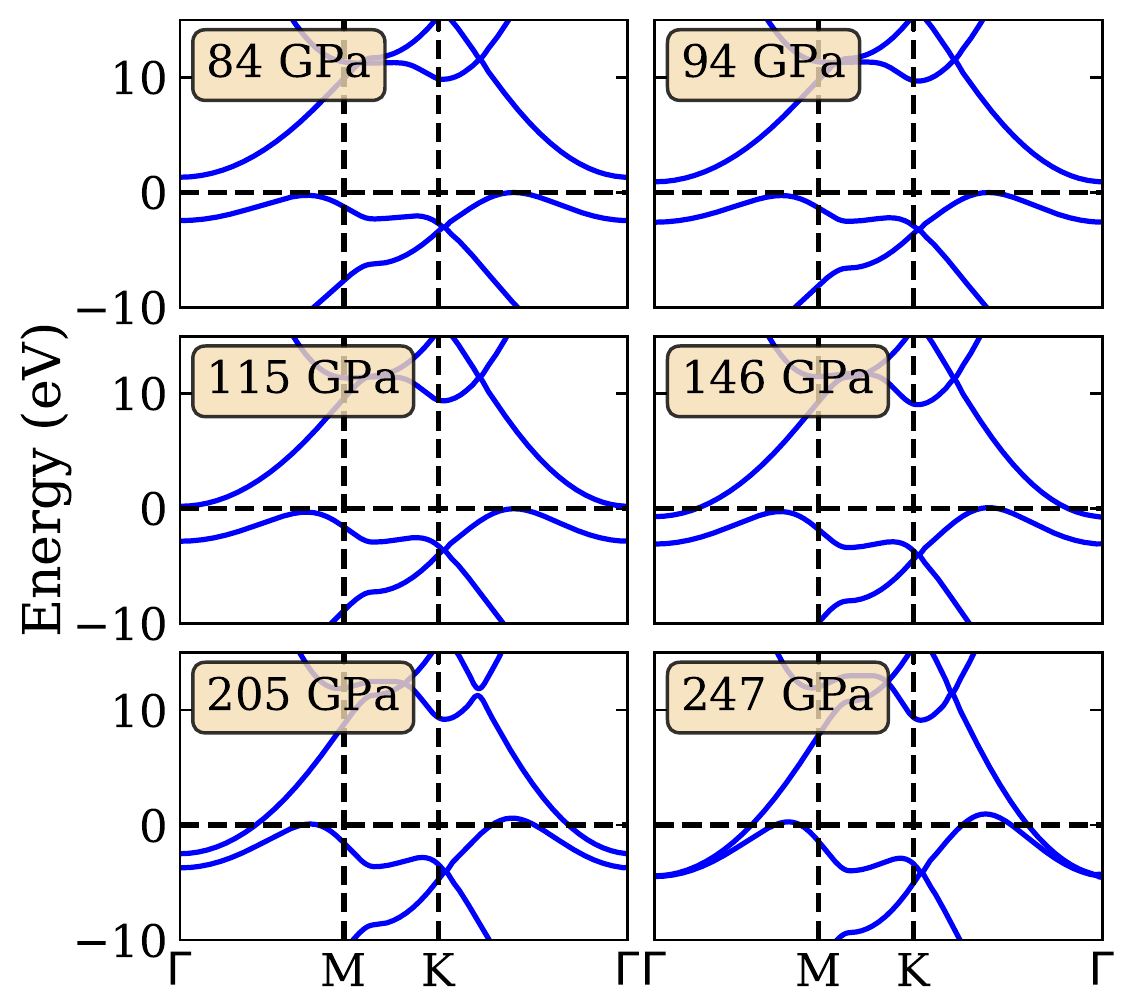}
\caption{\label{dft-band} The DFT-BLYP band structure of $hcp$ structure at different pressures.}
\end{figure}

According to the DFT-BLYP band structure (figure~\ref{dft-band}) the conduction band minimum takes place at the high-symmetry $\Gamma$-point. Whereas the valence band maximum occurs between $K$ and $\Gamma$ points. The DFT band structure shows also a second valence band peak at k-point between $\Gamma$ and $M$ points, which is (depending on the pressure) in between 0.2 and 0.6 eV below the valence band maximum.

\begin{figure}
\centering
\begin{tabular}{c}
\includegraphics[width=0.5\textwidth]{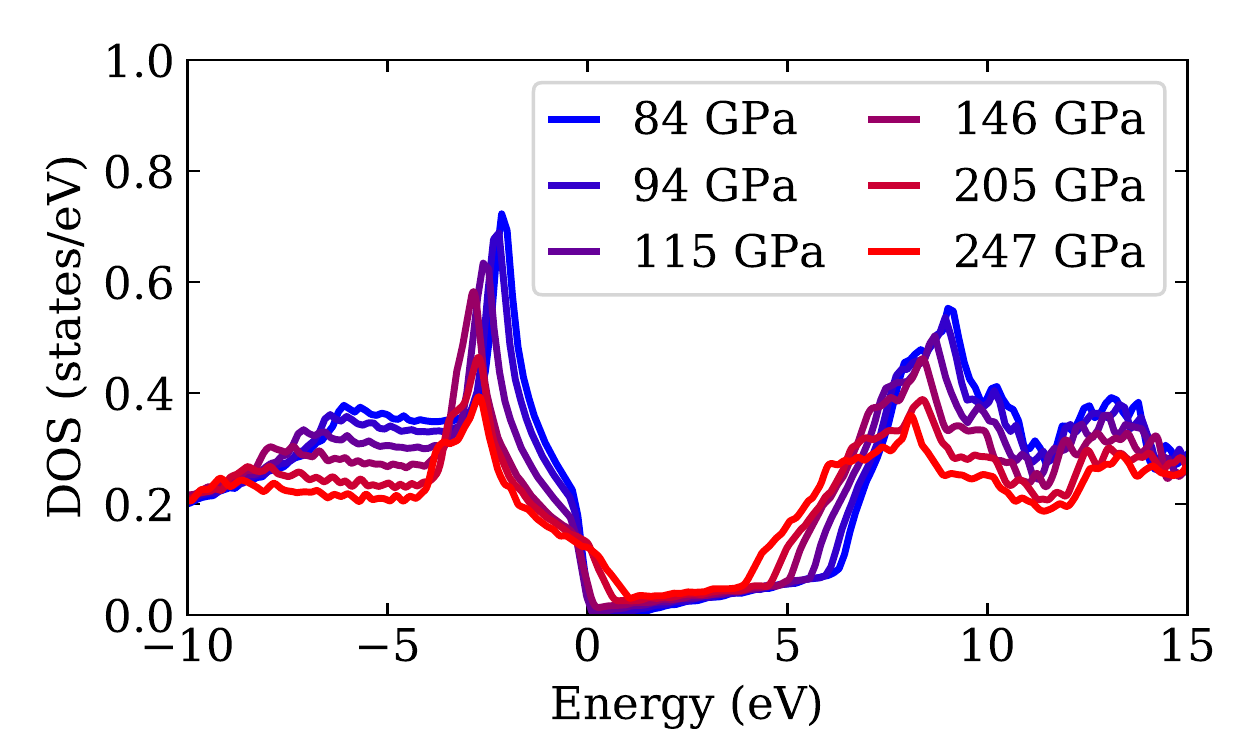} \\
\includegraphics[width=0.5\textwidth]{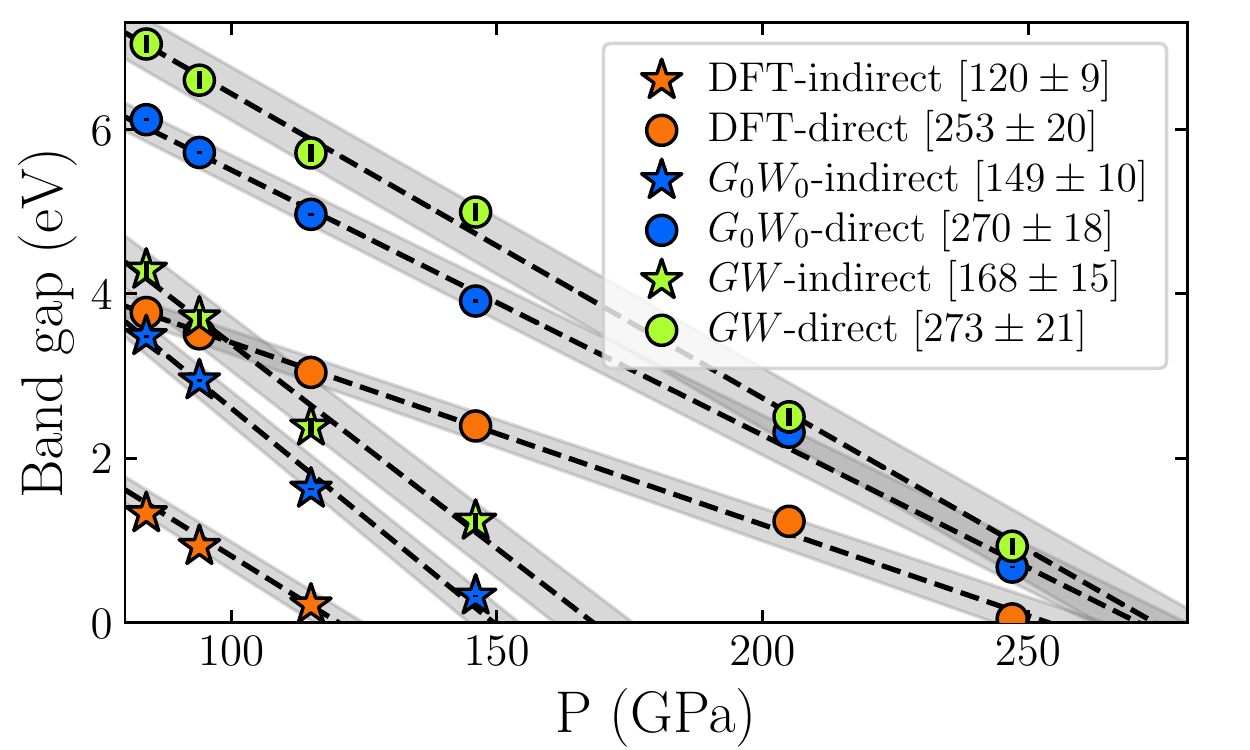} \\
\end{tabular}
\caption{\label{dos} (Top panel) The DFT-BLYP electronic density of states (DOS) of the $hcp$ structure as a function of pressure. (Bottom panel) DFT-BLYP, $\gowo$ and fully self-consistent $\gw$ energy band gaps as a function of pressure. Dashed lines are the linear fit of the corresponding data points explained in the legend, where corresponding gap closure pressures are specified as well. Shadowed regions covering the data points and extrapolation error bars are used to determine the gap closure pressure error bars.}
\end{figure}

Increasing the pressure results dominantly in shifts of the energy bands relative to the Fermi energy. It is true for both DFT and GW calculations. Therefore, the linear extrapolation to the zero band gaps was used for finding the associated pressures (Figure~\ref{dos}). Our DFT-BLYP calculations predict that the smallest indirect and direct band gaps close around 120 and 250 GPa, respectively. The $\ggf$ indirect and direct static (without protons zero point motion contribution) band gaps terminate at 150 and 270 GPa, respectively.  The fixed-node diffusion quantum Monte Carlo (DMC) calculation suggested that in majority of molecular structures of high-pressure solid hydrogen the indirect energy band gap reduction per pressure is $\sim 0.02$ eV/GPa~\cite{PRB17}. However, a linear fitting of $\ggf$ band gaps as function of pressure yielded 0.05 and 0.03 eV/GPa indirect and direct band gap reduction rates, respectively. 

For the sake of comparison, we performed also the fully self-consistent $\gw$ ($
\scgw$) calculations following the identical algorithms for QP energies extraction and the bandgaps extrapolation. The $
\scgw$ indirect and direct band gap closure points are at 170 and 273 GPa, respectively, and corresponding gap reduction rates are 0.05 and 0.037 eV/GPa. Since the self-consistent $\gw$ gaps are typically overestimated with respect to experiments, the resulting gap closure pressures should be seen as upper bounds.
%

The DFT-BLYP electronic density of states (DOS) (figure~\ref{dos}) illustrates that the indirect band gap closure has a minor contribution to the DOS at the Fermi level. Instead, an essential contribution originates from the hole-like bands crossing the Fermi level. Thus, it can be speculated that the system possibly behaves as a bad metal with properties similar to a semimetal at lower pressures, which has also been claimed by experiment~\cite{Eremets19}. This feature is almost independent of the crystal structure of solid molecular hydrogen. In most of the theoretical studies of the molecular structures of high-pressure solid hydrogen, it has been observed that their indirect band gap closes 150-200 GPa (120GPa in our 
$\gowo$ case) before the termination of the direct band gap~\cite{Gorelov}. 

So far, we have discussed the static band gap termination, meaning that the proton zero point motion (ZPM) contribution, which reduces the energy band gap, is neglected. The effect of ZPM and ionic thermal contribution to the fundamental band gap reduction, which is estimated  using the path-integral-molecular-dynamic~\cite{JCC} and Coupled Electron Ion Monte Carlo~\cite{Gorelov}, is 2-2.5~eV depending on the crystal structure of solid hydrogen and the pressure range. Recently DFT-BLYP method is applied on a DFT-PBE predicted molecular structure and the results predict that nuclear quantum fluctuations can reduce the optical band gap by 3.0 eV and indirect gap by 2.0 eV~\cite{Monacelli}. Assuming the band gap reduction being larger than $2$ eV in both gaps due to ZPM, one can estimate that the $\ggf+$ZPM direct band gap terminates below $\sim210$ GPa, while the indirect one should close below $\sim110$ GPa. Predicted direct band gap closure pressure is below the experimental~\cite{CJi} value of at least $250$ GPa, which could be to either due to other non-accounted contributions or due to possible dependence of the ZPM correction magnitude on pressure. Since this dependence is unknown, we take the predicted pure $\ggf$ bandgap closure pressure of 270 GPa as the upper bound.

Most of the experimental studies suggest that the metallic behaviour of solid hydrogen arises only at pressures larger than $\sim 350$ GPa. For instance, measurements by Eremets $\it {et~al.}$ predict that at pressure of 350-360 GPa and $T<200 K$, the hydrogen starts to conduct with a semimetallic behaviour~\cite{Eremets19}. Another report of recent experiment \citep{Loubeyre20} yield a discontinuous change of the direct band gap of high-pressure hydrogen, from 0.6 eV to below 0.1 eV at $\sim$425 GPa. The results were associated with the formation of metallic state for the high-pressure hydrogen. Hence, the structure of high-pressure solid hydrogen has to have a finite energy band gap at pressures up to 350 GPa, which is not the case for the $hcp$ even without the ZPM contribution. We then conclude that the $hcp$ structure can not be the right candidate for the structure of phase III of solid hydrogen due to inconsistency between its $\ggf$ metallization pressure and experiment. 

\section{\label{con} Conclusion}
We presented and discussed our implementation of the GW space-time method which allows us to calculate the properties of extended systems at finite temperature. We have calculated QP-energies using two approaches. In the first one, named $\gac$, the analytic continuation of the self energy is employed. In our second approach $\ggf$, we obtained the valence- and conductance-band energies from the asymptotic decay of the Green’s function at long imaginary times. To validate our implementation, we calculated the energy band gap of two well-studied systems of crystalline Si and Ge, and compared our data with experiment and established benchmark. We found that Chebyshev representation of the imaginary time axis is efficient at controlling the corresponding systematic errors in the numerical representation of G and W. Our robust results with controlled systematic errors within the GW approximation, agree well with experiment and other published works. 

The GWST scheme based on the norm-conserving pseudopotentials allows to store G and W fully in the allocatable memory. Using basis sets, as in the most of the GW implementations, requires storing one of G or W as a four-dimensional (in basis index) array or include non-trivial bare vertex element into the diagrammatic rules or construct these quantities in the real space from an intermediate representation~\cite{Kutepov2009,Kutepov2012,Liu2016}. Having all quantities in the real space and stored in the dynamic memory allocation eliminates these problems and enables quick stochastic sampling of the higher-order diagrams by means of the established DiagMC methodology~\cite{Prokofiev98, Kozik, Houcke, Mishchenko,van2012feynman,deng2015emergent,Chen19}. DiagMC allows to stochastically sum the series in terms of G and W in a numerically exact way, with or without self-consistency  which can make calculations of materials properties fully controlled and reliable in the future.

We applied our $\ggf$ technique to the problem of band gap closure of solid hydrogen. We calculated the band structure, direct and indirect band gaps of experimentally observed $hcp$ solid hydrogen. Considering the linear behaviour of band gap with respect to pressure, the extrapolation of band gaps to zero indicates that the $\ggf$ direct and indirect band gaps terminate around 270 and 150 GPa, respectively. Taking into account zero point motion (ZPM) electronic band renormalization, we estimated that both $\gowo$ direct and indirect band gaps would be closed at pressures below $210$ GPa. 
Although our prediction contradicts with experiment~\cite{CJi}, where insulating $hcp$ structure was observed at $250$ GPa, there might be other forces, which increase the band gap and band gap closure pressure directly or decrease the ZPM correction with pressure. Assuming the last scenario, and noting that ZPM always reduces the bandgap, we can take the $\ggf$ result with $270$ GPa band gap closure pressure as an upper pressure limit for the insulating $hcp$ phase.
Most of experimental data, however, indicate that the solid hydrogen energy band gap at T $<$ 250 K remains open up to $\sim 350$ GPa. Thus, the structure of solid hydrogen at pressures below $350$ GPa has to have a non-zero energy band gap. 
The comparison of our results with this experimental data  rules out the possibility of realisation of a $hcp$ structure in the phase III of high-pressure solid hydrogen, unless the GW approximation is dramatically inadequate for this problem, which deserves a separate investigation in the future. The small discrepancy between the $\gowo$ and the self-consistent GW results indicates that such a substantial correction, if any at all, should come from vertex corrections. The prediction of instability of the $hcp$ phase is inconsistent with the recent X-ray diffraction measurements~\cite{CJi}. According to energy band gap analysis and the suggestion by experiment that the metallization of hydrogen proceeds within the molecular solid~\cite{Loubeyre20}, the likely candidate for the structure of phase III of solid hydrogen should have a finite static (ignoring the zero point motion) direct energy band gap at $\sim400$ GPa. 

In most of the DFT-predicted molecular structures for the phase III, the fundamental energy band gap is indirect. By increasing the pressure, indirect band gap closure takes place before direct band gap and the metallization or semi-metallization occurs through indirect band gap termination.  Correspondingly, some of the other PBE-DFT predicted candidates for the phase III with indirect band gap at pressures above 300 GPa have symmetries of $C2/c$, $Cmca$, $Pc$, and $Pbcn$ with 24 (or 12), 24 (or 12), 48, and 48 number of hydrogen atoms per primitive unit cell. These structures are more complicated than $hcp$ and deserve further investigation. Our approach allows to systematically study them. 

\section{Acknowledgement}
This work was supported by the Simons Foundation as part of the Simons Collaboration on the Many-Electron Problem. A.D. was "also" supported by "the NCCR MARVEL, funded by the Swiss National Science Foundation". 

\newpage

\bibliography{main}

\end{document}